# Robust scatter correction method for cone-beam CT using an interlacing-slit plate[*]

HUANG Kui-Dong(黄魁东)[1], XU Zhe(徐哲)[1], ZHANG Ding-Hua(张定华)[1], ZHANG Hua(张华)[2], SHI Wen-Long (时稳龙)[1]

[1] Key Lab of Contemporary Design and Integrated Manufacturing Technology of the Ministry of Education, Northwestern Polytechnical University, Xi'an 710072, China
[2] School of Mechanical Engineering, Northwestern Polytechnical University, Xi'an 710072, China

**Abstract:** Cone-beam computed tomography (CBCT) has been widely used in medical imaging and industrial nondestructive testing, but the presence of scattered radiation will cause significant reduction of image quality. In this article, a robust scatter correction method for CBCT using an interlacing-slit plate (ISP) is carried out for convenient practice. Firstly, a Gaussian filtering method is proposed to compensate the missing data of the inner scatter image, and simultaneously avoid too-large values of calculated inner scatter and smooth the inner scatter field. Secondly, an interlacing-slit scan without detector gain correction is carried out to enhance the practicality and convenience of the scatter correction method. Finally, a denoising step for scatter-corrected projection images is added in the process flow to control the noise amplification. The experimental results show that the improved method can not only make the scatter correction more robust and convenient, but also achieve a good quality of scatter-corrected slice images.

## 1. Introduction

Cone-beam computed tomography (CBCT) is one of the hot spots in the fields of medical imaging and industrial nondestructive testing, and there have been some considerable advances over the past ten years [1-3]. Compared with fan-beam CT, CBCT scans the object with area array detectors, mostly flat panel detectors (FPDs), and reconstructs a volume data for the detection and analysis, giving higher scanning efficiency and radiation utilization [4, 5].

Scatter is an important factor affecting the quality of CT slice images. It mainly reduces image contrast and blurs image details. The goal of scatter correction is to make the reconstructed images reflect the detected object's real information by reducing or eliminating the adverse effects caused by scatter through some appropriate methods. In general, the existing scatter suppression and correction methods can be divided into three types: hardware-based, software-based, and hybrid. Hardware-based correction methods add correction tools to CT systems to reduce the scatter rays obtained by the detectors; these tools include air-gaps, filter plates, collimators, compensators [6], cellular or linear radiopaque grids [7], and so on. Software-based correction methods analyze the properties of detected objects and projection images, and then process the data by digital image processing methods to get the scatter images used to correct scatter influences; these methods include convolution [8], deconvolution, Monte Carlo simulation [9, 10], and so on. Hybrid correction methods not only add some correction tools in the CT system, but also estimate scatter distributions using algorithms such as beam-stop array (BSA) [11], beam-hole array (BHA) [12], beam attenuation array [13], stationary beam blocker [14], moving blocker strips [15], rotating blocker strips [16], attenuation baffle [17], primary ray modulation [18], and so on.

Recently we proposed a scatter correction method for CBCT based on an interlacing-slit scan [19]. In this method, we designed and manufactured an interlacing-slit plate (ISP) with interlaced slits. The scatter-suppressed projection images can be obtained directly with ISP scans, and then inner-scatter-corrected projection images are calculated out based on image fusion. The method needs two whole scans, and the distance between ray source and ISP cannot be changed greatly. Compared with other popular methods, especially BSA and BHA, our approach is essentially a collimator method. It has fewer assumptions and approximations, and its distinguishing features are no object scatter computation, no cone angle effect, and the inclusion of inner scatter correction. In theory, if the interlacing slits are thick and dense enough, the scatter correction result with ISP in CBCT can be close to the scatter suppression result with collimator in fan-beam CT.

In this article, we improve the reliability and convenience of this method from three aspects, including more accurate and reliable calculation for the inner scatter field, interlacing-slit scan without FPD gain correction, and noise control and evaluation. The feasibility and practicality of the improvements were verified by experiments.

[*] Supported by the National Science and Technology Major Project of the Ministry of Industry and Information Technology of China (Grant No. 2012ZX04007021), the Aeronautical Science Fund of China (Grant No. 2014ZE53059), and the Fundamental Research Funds for the Central Universities of China (Grant No. 3102014KYJD022).
E-mail: kdhuang@nwpu.edu.cn

## 2. Materials and Methods

### 2.1 Inner scatter field calculation

The structure of an ISP is shown in Fig. 1(a), where the slits in the regions of A and B are interlacing. Fig. 1(b) is the diagram of a scan with slits A of the ISP. According to our previous study, there are apparent outputs in the blocker strip regions of ISP, where the strip is too thick to be irradiated by the penetrating rays. So scatter existing there in the FPD cannot be caused by the detected object. We called it the inner scatter field here. In some related investigations it is also called veiling glare [20, 21]. In addition, the inner scatter distribution is not average. It is related to the distribution of the received radiation from adjacent slit regions.

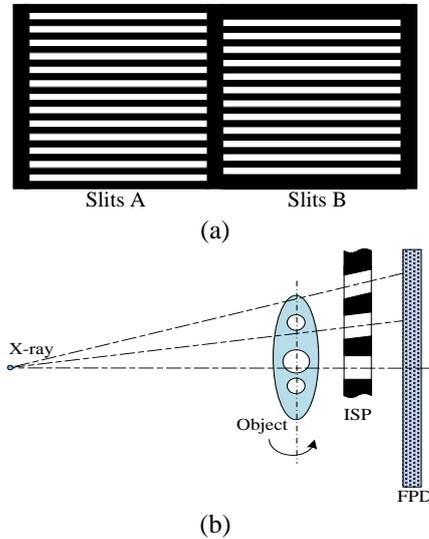

Fig. 1. Diagrams of the scan with ISP. (a) Structures of ISP, (b) Scan with slits A of ISP.

When calculating the inner scatter field, we cannot get a complete image by summing up the segmented blocker strip regions of slits A and slits B in the scans with ISP, because the width of the blocker strips is smaller than that of the slits. We can only get an image with some small strips, where the gray values are zero. In the previous method, to compensate for the missing data, we calculate the sum of gray values in the pixel's neighborhood whose grayscale is zero, and the number of non-zero pixels in this neighborhood, then the pixel's grayscale is equal to the quotient of dividing the sum by the quantity. In practice, we found that in slit regions with small gray values, the inner scatter calculated using this method may be relatively large, which causes the pixel grayscale to be too small after the inner scatter correction. Sometimes this will result in serious noise in reconstructed slice images.

In this article, we note that the distribution of inner scatter should be smooth and the grayscale near a blocker strip's edge is usually higher than that in the strip's interior. Then we use Gaussian filtering to compensate the missing data in the inner scatter image. In general, the filtering window should not be less than 2 times the width of the missing data area. The filtering strength is usually equal to the standard deviation of the blocker strip regions. This method can solve the problem that the calculated inner scatter may be too large, which makes the corrected gray values close to zero. In addition, the calculated inner scatter field with filtering can be smoother.

### 2.2 Interlacing-slit scan without FPD gain correction

In the previous method, we needed to collect four groups of projection images which have been processed with some FPD corrections. One of the important corrections is gain correction. In general, if the CBCT scanning voltage is changed we need to redo the gain correction. That means we need to completely move the ISP away from the FPD to get the gain coefficient image. We notice that the way of obtaining the fusion coefficient image in the previous method is similar to the operation of getting the gain coefficient image. So, we propose an interlacing-slit scan method without FPD gain correction, which can simplify the scanning steps.

Firstly, we will analyze the calculation process of the previous method shown in Fig. 2(a), where $T(x,y)$ is the penetration value through the slits, $S(x,y)$ is the inner scatter value, $\mu_1(x,y)$ is the gain coefficient, $\mu_2(x,y)$ is the fusion coefficient, and $(x,y)$ is the pixel location in the FPD. $\mu_1(x,y)$ is usually calculated as

$$\mu_1(x,y) = \frac{\overline{T(x,y)+S(x,y)}}{T(x,y)+S(x,y)} \quad (1)$$

where '$-$' represents the average of the whole image. After the inner scatter correction, we can get $T'(x,y)$, and $T'(x,y) = \mu_1(x,y) \cdot T(x,y)$. Then the calculation of $\mu_2(x,y)$ can be described as

$$\mu_2(x,y) = \frac{\overline{T'(x,y)}}{T'(x,y)} \quad (2)$$

After analyzing and comparing the two formulas above, we find that their calculation models are very similar. So, as shown in Fig. 2(b), we combine the gain correction and image fusion into one step in this article. Equation (3) describes the calculation of the new fusion coefficient $\mu(x,y)$.

$$\mu(x,y) = \frac{\overline{T(x,y)}}{T(x,y)} \quad (3)$$

Comparing the previous method with the proposed method, the deviation $E$ of the calculations is shown in Equation (4).

$$\begin{aligned} E &= \mu(x,y)T(x,y) - \mu_1(x,y)\mu_2(x,y)T(x,y) \\ &= [\mu(x,y) - \mu_1(x,y)\mu_2(x,y)]T(x,y) \\ &= [\frac{\overline{T(x,y)}}{T(x,y)} - \frac{\overline{T(x,y)+S(x,y)}}{T(x,y)+S(x,y)} \cdot \frac{\overline{T'(x,y)}}{T'(x,y)}]T(x,y) \\ &= [\frac{\overline{T(x,y)}}{T(x,y)} - \frac{\overline{T(x,y)+S(x,y)}}{T(x,y)+S(x,y)} \\ &\quad \cdot \frac{\overline{T'(x,y)}}{\frac{\overline{T(x,y)+S(x,y)}}{T(x,y)+S(x,y)} \cdot T(x,y)}]T(x,y) \\ &= \overline{T(x,y)} - \overline{T'(x,y)} \end{aligned} \quad (4)$$

The value of $E$ is usually very small and can be omitted. So, the proposed calculation is approximately equal to the previous method.

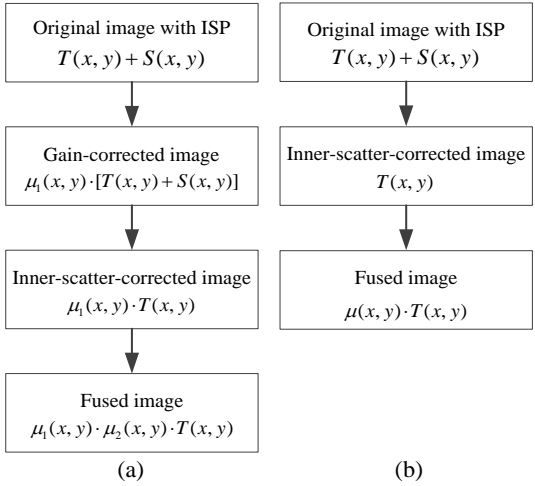

Fig. 2. Comparison of the calculation flows. (a) Previous method, (b) Proposed method.

**2.3 Noise control in scatter correction**

After the scatter suppression with slits and inner scatter correction, the grayscale of the scatter-corrected projection images will be significantly reduced. The noise level in CBCT systems is usually considered approximately unchanged if using the same scanning parameters. The total grayscale is decreased, and the proportion of noise is increased. So it is bound to make the signal to noise ratio (SNR) lower in slice images reconstructed with scatter-corrected projection images. In the previous method, we increased the exposure level to compensate the SNR and achieved a certain effect. However, due to the limitation of X-ray source power, it is difficult, even impossible, to increase the SNR to the level of no scatter correction.

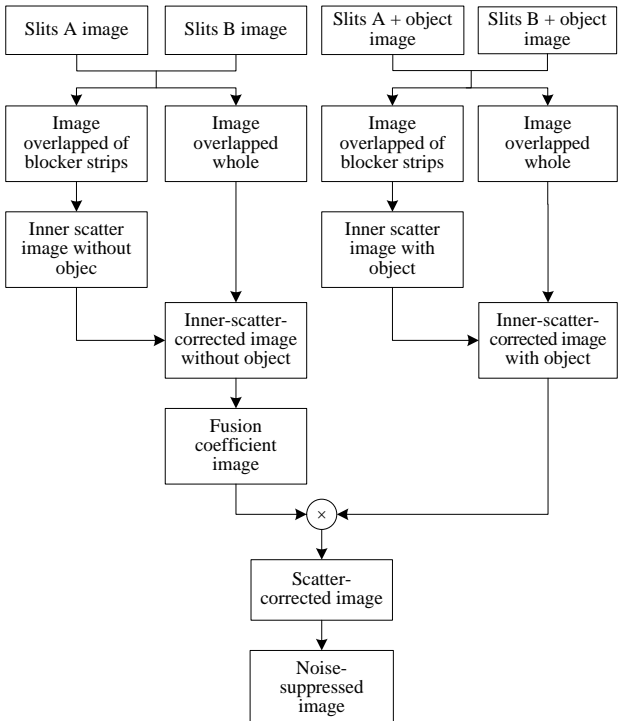

Fig. 3. Total process flow of the robust scatter correction method.

Since the main reason of SNR decrease or noise amplification is the relatively increased noise in projection images, we add a denoising step after the scatter correction for projection images in this article. Common denoising models include Gaussian filtering [22], anisotropic diffusion filtering [23], bilateral filtering [24], full variational filtering [25], wavelet transform filtering [26], and non-local mean filtering (NLM) [27]. We will compare the noise control effects with some typical methods in the subsequent experiment.

According to the analysis above, we can get the total process flow of the robust scatter correction for CBCT with interlacing-slit scan, as shown in Fig. 3.

## 3. Experiments and discussion

### 3.1 Experiment with an iron object

The X-ray source of the experimental CBCT system was a Y.TU 450-D02 from Yxlon, and the FPD as a PaxScan 2520 from Varian. For the ISP used, the slit width as 8 mm, the spacing of slit centers was 15 mm, the plate thickness was 40 mm, and the material was lead-antimony alloy. CBCT scans with the ISP were carried out for an iron object. The scan voltage was 350 kV, and the exposure was 0.056 mA·s. The number of projection images scanned without the object was 9, and scanned with the part was 360. The resolution of projection images and slice images was the same, 1024×1024.

Fig. 4 is the comparison of the inner scatter fields calculated with the previous and proposed method. Here we find that there is apparent scatter in the blocker strip regions of the ISP, and the scatter distribution with and without the object is significantly different. For the inner scatter fields obtained by the proposed method, both with and without the object, they are obviously smoother than those obtained with the previous method. From Fig. 4(k), we find that the grayscale of the inner scatter image obtained by the proposed method is slightly smaller than that with the previous method, which can avoid the situation that some pixel values may be close to zero after the inner scatter correction.

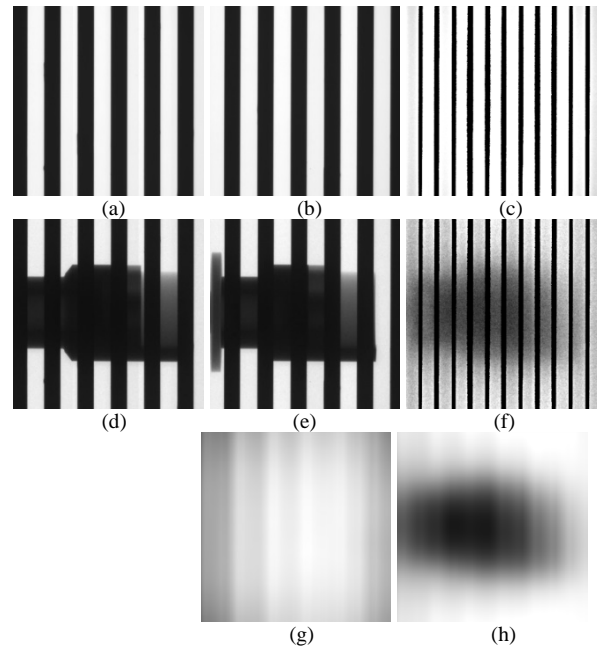

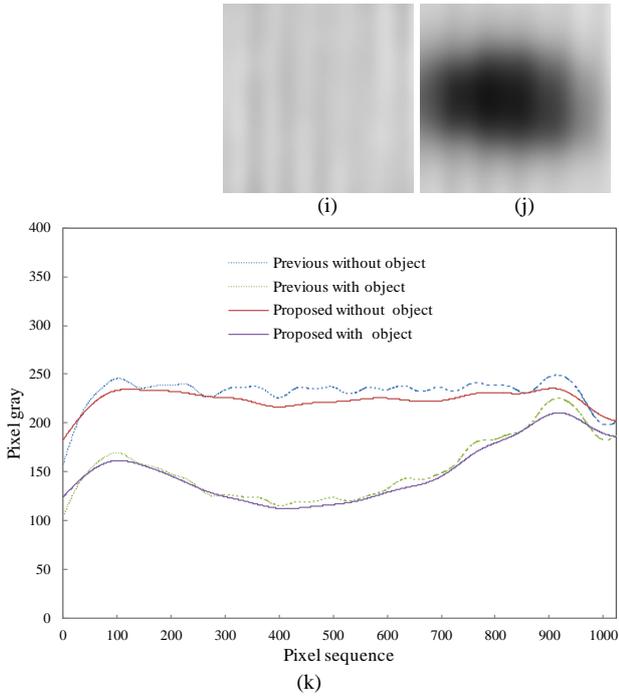

Fig. 4. Comparison of the inner scatter fields calculated with previous and proposed method. (a)-(c) are the images of slits A, slits B, and overlapped blocker strip regions of slits A and slits B, respectively, all without the object. (d)-(f) are the same as (a)-(c), but all with the object. (g) and (h) are the inner scatter field calculated with the previous method, without and with the object respectively. (i) and (j) are same as (g) and (h) but with the proposed method. (k) is the profile comparison on the horizontal position in the middle of (g)-(j). Display window: (a), (b), (d) and (e) are [0, 2600], the others are [120, 300].

Fig. 5 is the comparison of scatter correction results of the $180^{th}$ projection image. From Fig. 5(b) and (e), we find that the proposed method involves no FPD gain correction also obtains a normal image only by image fusion. Furthermore, Fig. 5(f) shows some details and differences. The gray values of the scatter-corrected images are both distinctly lower than that of the open-scan image. The previous method gives lower values, because its inner scatter is bigger. After further observation we can see that in the projection region of the thick part, some gray values calculated by the previous method are close to zero, which may cause serious noise in slice images, but the proposed method has avoided this situation.

In order to assess the noise control effects, we choose three typical filtering methods, Gaussian filtering, bilateral filtering and NLM filtering, to process the scatter-corrected projection images respectively. Three slice images reconstructed by the FDK algorithm [28] are selected to show the results, where different structures are included. Fig. 6, Fig. 7 and Fig. 8 show comparisons of the $285^{th}$, $580^{th}$ and $723^{th}$ slice image respectively. In the (a) of these three figures, line A is the profile location, rectangle B is the computing location of contrast to noise ratio (CNR) [29] and average gradient (AG) [30], and rectangle C is the computing location of SNR. The assessment results are shown in Table 1.

From the three figures we can see that the scatter-corrected slice images in (b) and (c) have obviously higher edge contrast than that of the open-scan in (a), and the noise level of the proposed method here is slightly lower than that of the previous method. For the noise control, all three filtering methods can reduce the noise level of slice images, but bilateral filtering and NLM filtering have better visual results.

A further more accurate comparison can be seen in Table 1. SNRs of the scatter-corrected slice images decrease significantly, while the other two evaluation items, CNRs and AGs, both increase obviously. After denoising, SNRs and CNRs can be both improved sharply. However, AGs are decreased seriously. The reason is the AG cannot distinguish noise and details well. So when the noise level is very high, AG is unrealistically high too, but after denoising, its value will decrease sharply. After the further comparison, we find that the SNR and CNR of NLM filtering are both higher than that of the other two, which means that NLM filtering can not only filter out most of the noise, but also keep the object details well. So NLM filtering has better results overall in this experiment.

The current denoising algorithms cannot entirely avoid the loss of image detail, so when using denoising algorithms to control the noise amplification after scatter correction, balance needs to be maintained between denoising and detail-keeping. In general, controlling the SNRs of scatter-corrected slice images to the level of the open-scan images would be good.

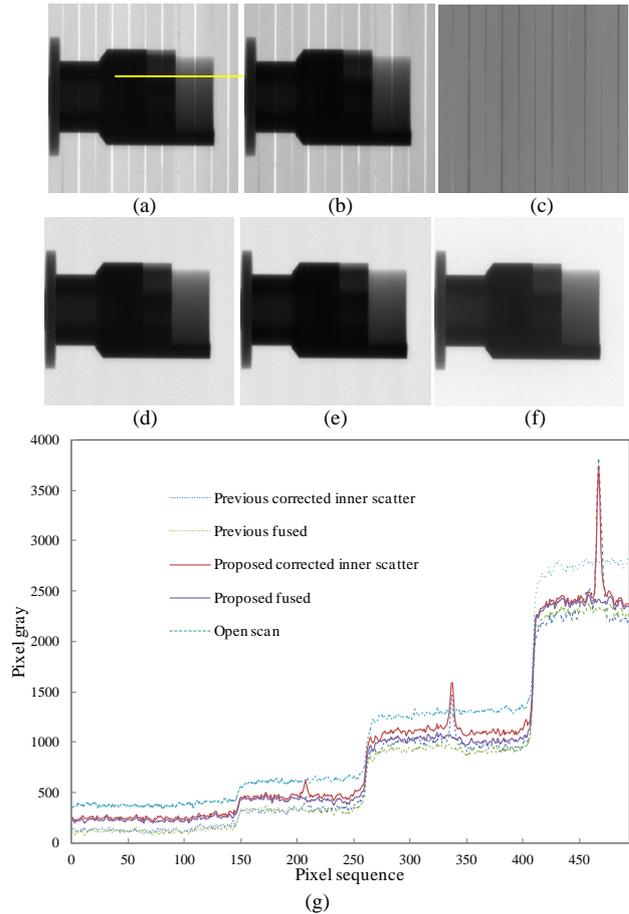

Fig. 5. Comparison of scatter correction results of the $180^{th}$ projection image. (a) and (b) are the inner-scatter-corrected image obtained with previous and proposed method, respectively. (c) is the fusion coefficient image. (d) and (e) are the fused image corresponding to (a) and (b), respectively. (f) is the open-scan image without ISP. (g) is the profile comparison on the horizontal line in (a). Display window: (c) is [-0.278, 2.30], the others are [0, 2600].

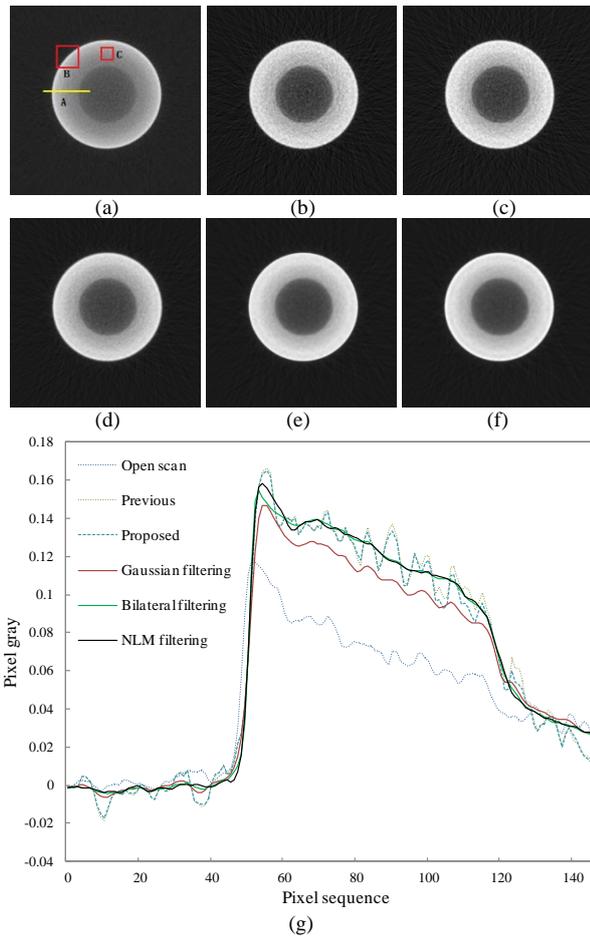

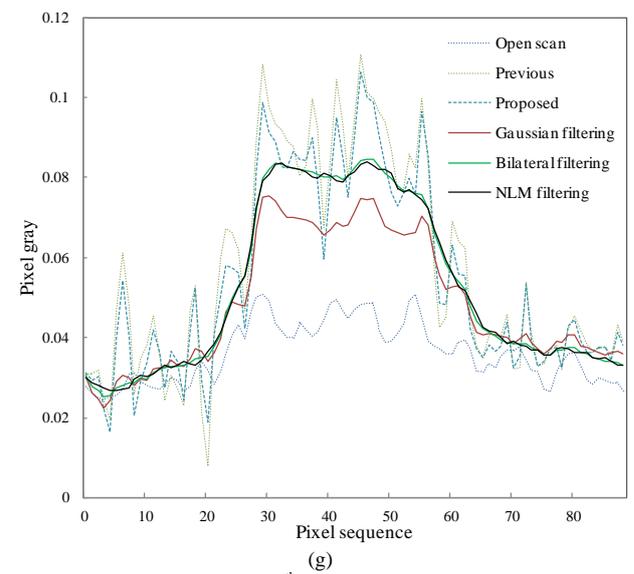

Fig. 7. Comparison of the 580th slice images before and after scatter correction. (a)-(c) are the slice image of open-scan, previous and proposed method without projection image denoising, respectively. (d)-(f) are the slice image of proposed method with projection images denoising by Gaussian filtering, bilateral filtering and NLM filtering, respectively. (g) is the profile comparison. Display window: [-0.03, 0.16].

Fig. 6. Comparison of the 285th slice images before and after scatter correction. (a)-(c) are the slice image of open-scan, previous and proposed method without projection image denoising, respectively. (d)-(f) are the slice image of proposed method with projection images denoising by Gaussian filtering, bilateral filtering and NLM filtering, respectively. (g) is the profile comparison. Display window: [-0.02, 0.16].

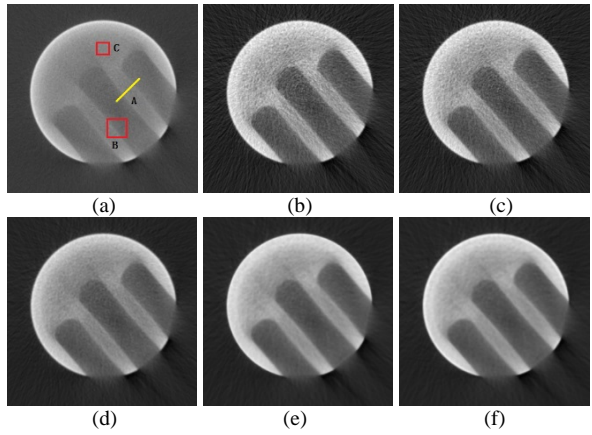

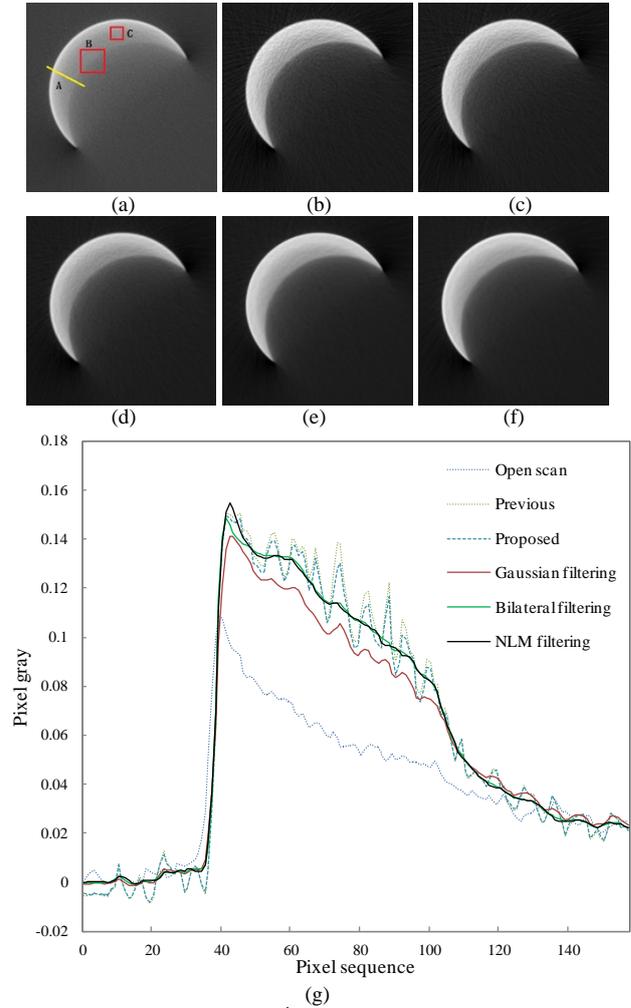

Fig. 8. Comparison of the 723th slice images before and after scatter correction. (a)-(c) are the slice image of open-scan, previous and proposed method without projection image denoising, respectively. (d)-(f) are the slice image of proposed method with projection images denoising by Gaussian filtering, bilateral filtering and NLM filtering, respectively. (g) is the profile comparison. Display window: [-0.02, 0.17].

Table 1. Evaluation comparison of three slice images with different methods.

| Slice | Method | SNR Value | SNR Improved (%) | CNR Value | CNR Improved (%) | AG Value | AG Improved (%) |
|---|---|---|---|---|---|---|---|
| 285th | Open scan | 14.3 | / | 6.30 | / | 0.00424 | / |
| | Previous | 10.7 | -25.2 | 7.80 | 23.8 | 0.00817 | 92.7 |
| | Proposed | 10.8 | -24.5 | 7.91 | 25.6 | 0.00786 | 85.4 |
| | Gaussian | 14.5 | 1.40 | 7.45 | 18.3 | 0.00443 | 4.48 |
| | Bilateral | 15.8 | 10.5 | 8.35 | 32.5 | 0.00439 | 3.54 |
| | NLM | 15.8 | 10.5 | 8.40 | 33.3 | 0.00469 | 10.6 |
| 580th | Open scan | 16.8 | / | 3.65 | / | 0.00157 | / |
| | Previous | 6.94 | -58.7 | 4.10 | 12.3 | 0.00561 | 257 |
| | Proposed | 7.21 | -57.1 | 4.11 | 12.6 | 0.00501 | 219 |
| | Gaussian | 15.3 | -8.93 | 4.84 | 32.6 | 0.00164 | 4.46 |
| | Bilateral | 17.6 | 4.76 | 5.23 | 43.3 | 0.00143 | -8.92 |
| | NLM | 18.1 | 7.74 | 5.50 | 50.7 | 0.00143 | -8.92 |
| 723th | Open scan | 19.8 | / | 3.43 | / | 0.00163 | / |
| | Previous | 11.0 | -44.4 | 4.31 | 25.7 | 0.00542 | 233 |
| | Proposed | 12.3 | -37.9 | 4.27 | 24.5 | 0.00486 | 198 |
| | Gaussian | 19.3 | -2.53 | 4.75 | 38.5 | 0.00143 | -12.3 |
| | Bilateral | 23.7 | 19.7 | 4.88 | 42.3 | 0.00126 | -22.7 |
| | NLM | 24.4 | 23.2 | 4.91 | 43.1 | 0.00128 | -21.5 |

### 3.2 Experiment of spatial and density resolution

To further validate our proposed scatter correction method, to improve the spatial resolution and density resolution, and to verify the generality of the method, another CBCT scanning experiment was carried out for the CTP682 module inside the Catphan©700 phantom. In this experiment, the X-ray source was a MXR-451HP/11 from Comet, and the FPD was a XRD 1621 AN15 ES from PerkinElmer. For the matched ISP, the slit width was 3 mm, the spacing of slit centers was 5.4 mm, the plate thickness as 20 mm, and the material was lead-antimony alloy. The scan voltage was 140 kV, the exposure was 0.95 mA·s, and the number of projections was 720. The resolution of projection images and slice images was the same, 1360×1360.

Fig. 9 is a comparison of the slice images with and without the proposed scatter correction. It is very clear that the visibility and clarity of the structures after scatter correction have both been improved. Furthermore, by using the tungsten wire in the CTP682 module, the modulation transfer functions (MTF) were measured and are shown in Fig. 10. Numerical comparison of the MTFs is shown in Table 2. We can see that the spatial resolution has been significantly enhanced after the proposed scatter correction, and when the MTF takes 5%, the spatial resolution is 2.40 lp/mm.

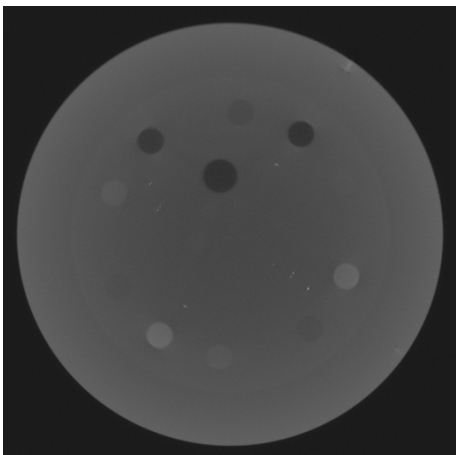

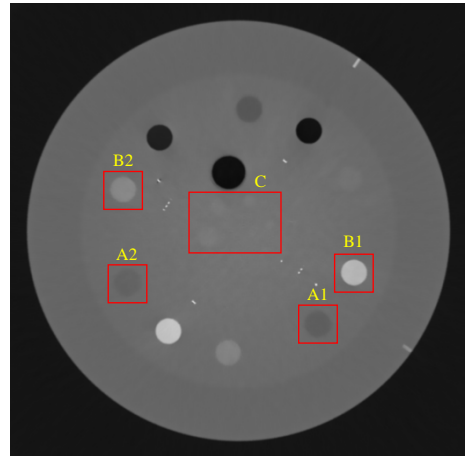

(b)
Fig. 9. Comparison of the slice images of the Catphan©700 with (a) open scan and (b) proposed scatter correction. Display window: [-0.008, 0.068].

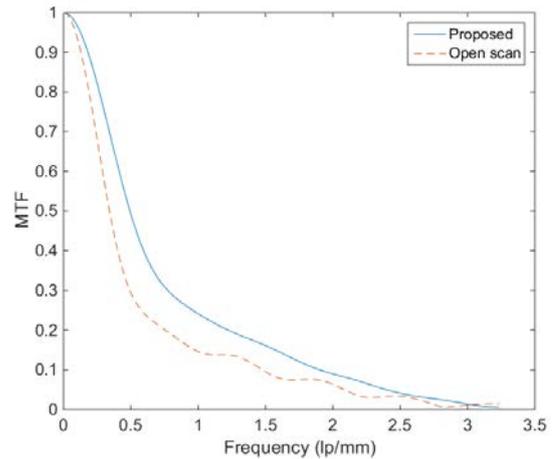

Fig. 10. Comparison of the MTFs of open scan and proposed scatter correction.

Table 2. Numerical comparison of the MTFs of open scan and proposed scatter correction.

| MTF | Open scan | Proposed | Improved (%) |
|---|---|---|---|
| 10% | 1.48 | 1.91 | 29.1 |
| 5% | 2.08 | 2.40 | 15.4 |

On the other hand, we can directly observe the improvement of density resolution after the proposed scatter correction from Fig. 9 (a) and Fig. 9 (b). In particular, some significant circular structures can be seen in region of interest (ROI) C in Fig. 9 (b), but in the same region in Fig. 9(a), there are no corresponding structures. For further comparative analysis, we chose two groups of different materials to calculate, each group comprising two materials with similar densities, as shown in Table 3, where $Z_{eff}$ is the effective atomic number, and the positions of the ROIs are marked in Fig. 9 (b). From the comparison results in Table 4, we can see that the contrasts and CNRs of the two groups both increase more than 100% after our scatter correction. At the same time, we also note that the contrast contribution of $Z_{eff}$ is significantly greater than that of density.

As we know, the scatter in CBCTs is associated with many factors, and there may be a big difference in the scatter correction results obtained under different experimental conditions. The qualities of slice images have been significantly improved with our scatter correction method on two CBCTs, which shows the proposed method has good robustness and practicality.

Table 3. Two groups of different materials.

| | ROI | Material (Circular Region) | $Z_{eff}$ | Density (g/cm$^3$) | Density difference(%) |
|---|---|---|---|---|---|
| Group A | A1 | Low density polyethylene | 5.44 | 0.92 | / |
| | A2 | Polystyrene | 5.70 | 1.03 | 12.0 |
| Group B | B1 | Bone 50% | 11.46 | 1.40 | / |
| | B2 | Delrin | 6.95 | 1.42 | 1.43 |

Table 4. Comparison results of the two groups on open scan and proposed scatter correction.

| | ROI | Mean (Circular Region) | \|Contrast\| | Improved (%) | CNR | Improved (%) |
|---|---|---|---|---|---|---|
| Open scan | A1 | 0.0151 | 0.0004 | / | 2.31 | / |
| | A2 | 0.0155 | | | 1.92 | / |
| | B1 | 0.0213 | 0.0032 | / | 4.73 | / |
| | B2 | 0.0181 | | | 3.10 | / |
| Proposed | A1 | 0.0201 | 0.0015 | 275 | 5.82 | 152 |
| | A2 | 0.0216 | | | 4.41 | 130 |
| | B1 | 0.0462 | 0.0137 | 328 | 11.6 | 145 |
| | B2 | 0.0325 | | | 7.31 | 136 |

## 4. Conclusions

In this article, we proposed a robust scatter correction method for CBCT using an ISP. The improvements mainly include three aspects. First, the use of Gaussian filtering to compensate the missing data of the inner scatter image, and simultaneously avoid too-large values of calculated inner scatter while smoothing the inner scatter field. Second, simplifying the previous interlacing-slit scan into one without detector gain correction, which can enhance the practicality and convenience of the scatter correction method. Third, adding a denoising step for scatter-corrected projection images in the process flow to control the noise amplification.

The experimental results show that the proposed method can not only make the scatter correction more robust and convenient, but also achieve a good quality of scatter-corrected slice images. These improvements can play an important role in preparing the interlacing-slit scan method for use in real CBCT systems.

## References

[1] Kruth J P, Bartscher M, Carmignato S, et al. CIRP Ann. - Manuf. Technol., 2011, **60**: 821-842
[2] Chiffre L D, Carmignato S, Kruth J P, et al. CIRP Ann. - Manuf. Technol., 2014, **63**: 655-677
[3] Lifton J J, Malcolm A A, McBride J W. Meas. Sci. Technol., 2015, **26**: 035003
[4] Endo M, Mori S, Tsunoo T, et al. IEEE Trans. Nucl. Sci., 2003, **50**: 1667-1671
[5] WANG X, YAN B, LIU H, et al. Acta Phys. Sin., 2013, **62**: 098702
[6] Graham S A, Moseley D J, Siewerdsen J H, et al. Med. Phys., 2007, **34**: 2691-2703
[7] Coatesa A M, Dixonb A M, Huckleb J. Radiography, 2012, **18**: 250-255
[8] LI Q, YAN B, LI L, et al. High Power Laser and Particle Beams, 2012, **24**: 2235-2238
[9] Mettivier G, Lanconelli N, Lo M S, et al. IEEE Trans. Nucl. Sci., 2012, **59**: 2008-2019
[10] Baer M, Kachelrieß M. Phys. Med. Biol., 2012, **57**: 6849-6867
[11] Ning R, Tang X, Conover D. Med. Phys., 2004, **31**: 1195-1202
[12] Schörner K, Goldammer M, Stephan J. Nucl. Instrum. Methods Phys. Res. Sect. B-Beam Interact. Mater. Atoms, 2011, **269**: 292-299
[13] ZHANG D H, HU D C, HUANG K D, et al. China Mechanical Engineering, 2009, **20**: 639-643
[14] NIU T, ZHU L. Med. Phys., 2011, **38**: 6027-6038
[15] Ouyang L, Song K, Wang J. Med. Phys., 2013, **40**: 071903
[16] XIE S P, LUO L M. Chin. Phys. C, 2012, **36**: 566-572
[17] XIE S P, LUO L M. Acta Electronica Sinica, 2011, **39**: 1708-1711
[18] Gao H, Fahrig R, Bennett N R, et al. Med. Phys., 2010, **37**: 934-946
[19] HUANG K D, ZHANG H, SHI Y K, et al. Chin. Phys. B, 2014, **23**: 098106
[20] Seibert J A, Nalcioglu O, Roeck W W. Med. Phys., 1984, **11**: 172-179
[21] Lazos D, Williamson J F. Med. Phys., 2012, **39**: 5639
[22] Babaud J, Witkin A P, Baudin M, et al. IEEE Trans. Pattern Anal. Mach. Intell., 1986, **8**: 26-33
[23] Alvarez L, Lions P L, Morel J M. SIAM J. Numer. Anal., 1992, **29**: 845-866
[24] Elad M. IEEE Trans. Image Process., 2002, **11**: 1141-1151
[25] Rudin L I, Osher S, Fatemi E. Physica D: Nonlinear Phenomena, 1992, **60**: 259-268
[26] Sharma P, Khan K, Ahmad K. Int. J. Wavelets Multiresolut Inf. Process., 2014, **12**: 1450038
[27] Buades A, Coll B, Morel J M. Multiscale Model. Simul., 2005, **4**: 490-530
[28] Feldkamp L A, Davis L C, Kress J W. J. Opt. Soc. Am. A, 1984, **1**: 612-619
[29] ZHANG H, HUANG K D, SHI Y K, et al. Computerized tomography theory and applications, 2012, **21**: 247-254
[30] HUANG K D, ZHANG D H, LI M J, et al. Acta Phys. Sin., 2013, **62**: 210702